\let\ifarxiv=\iffalse    % JOURNAL VERSION
\documentclass[12pt]{scrartcl}
\ifarxiv\ifnum\pdfoutput=1\else
\PassOptionsToPackage{hypertex}{hyperref}
\PassOptionsToPackage{draft}{graphicx}
\fi\fi

\usepackage{amsmath,amssymb}
\usepackage[bookmarks=true,hyperfigures=true]{hyperref}
\usepackage{graphicx}
\usepackage[nosort]{cite}
\usepackage[bulletsep]{collref}
\usepackage{feynmp}
\usepackage{multirow}
\usepackage{listings}
\usepackage{longtable}
\usepackage{multicol}
\usepackage{hhline}
\usepackage{pslatex}

\graphicspath{{figs/}}
\usepackage{color}
\usepackage{colordvi}
\usepackage{color}
\usepackage{colordvi}

\definecolor{mygreen}{rgb}{0,0.75,0}

\allowdisplaybreaks[3]

\usepackage[font=small,labelfont=bf,width=0.85\textwidth]{caption}

\makeatletter
\let\old@startsection=\@startsection
\renewcommand{\@startsection}[6]{\old@startsection{#1}{#2}{#3}{#4}{#5}{#6\mathversion{bold}}}
\makeatother

\makeatletter
\newlength{\apb@width}
\newcommand{\autoparbox}[2][c]{\settowidth{\apb@width}{#2}\parbox[#1]{\apb@width}{#2}}

\makeatother

\let\oldPhi=\Phi
\let\oldPsi=\Psi
\let\oldGamma=\Gamma
\let\oldDelta=\Delta
\let\oldSigma=\Sigma
\let\oldLambda=\Lambda
\let\oldTheta=\Theta
\let\oldPi=\Pi
\let\oldXi=\Xi
\let\oldUpsilon=\Upsilon
\let\oldOmega=\Omega
\renewcommand{\Phi}{\mathnormal{\oldPhi}}
\renewcommand{\Psi}{\mathnormal{\oldPsi}}
\renewcommand{\Gamma}{\mathnormal{\oldGamma}}
\renewcommand{\Sigma}{\mathnormal{\oldSigma}}
\renewcommand{\Delta}{\mathnormal{\oldDelta}}
\renewcommand{\Theta}{\mathnormal{\oldTheta}}
\renewcommand{\Lambda}{\mathnormal{\oldLambda}}
\renewcommand{\Pi}{\mathnormal{\oldPi}}
\renewcommand{\Xi}{\mathnormal{\oldXi}}
\renewcommand{\Upsilon}{\mathnormal{\oldUpsilon}}
\renewcommand{\Omega}{\mathnormal{\oldOmega}}

\ifx\genfrac\sdflkaj\else\fi
\newcommand{\sfrac}[2]{{\textstyle\frac{#1}{#2}}}

\newcommand{\vev}[1]{\langle#1\rangle}
\newcommand{\bev}[1]{ [#1]}
\newcommand{\pslash}{{p\!\!\!\!/}}

\newcommand{\be}{\begin{equation}}
\newcommand{\ee}{\end{equation}}

\newcommand{\Tr}{\mathop{\mathrm{Tr}}}

\makeatletter
\def\mr@ignsp#1 {\ifx\:#1\@empty\else #1\expandafter\mr@ignsp\fi}%
\newcommand{\multiref}[1]{\begingroup%\let\protect\string%
\xdef\mr@no@sparg{\expandafter\mr@ignsp#1 \: }%
\def\mr@comma{}%
\@for\mr@refs:=\mr@no@sparg\do{\mr@comma\def\mr@comma{,}\ref{\mr@refs}}%
\endgroup}
\makeatother

\ifx\href\asklfhas\newcommand{\href}[2]{#2}\fi

\newcommand{\hypref}[2]{\ifx\href\asklfhas #2\else\href{#1}{#2}\fi}
\newcommand{\Secref}[1]{Section~\multiref{#1}}

\newcommand{\figref}[1]{Fig.~\multiref{#1}}
\renewcommand{\eqref}[1]{(\multiref{#1})}
\def\Eqn#1{Eq.~\eqref{#1}}

\def\ib{{\bar\imath}}
\def\tree{{\rm tree}}

\newcommand{\indexfett}[2][f]{\if#1f{\index{#2|textbf}}\else{\index{#2}}\fi}

\ifx\hypersetup\sadfkjashdfkxja\newcommand\hypersetup[1]{}\fi

\hypersetup{plainpages=false}
\hypersetup{pdfpagemode=UseNone}
\hypersetup{bookmarksnumbered=true}
\hypersetup{pdfstartview=FitH}
\hypersetup{colorlinks=false}
\hypersetup{citebordercolor={.5 1 .5}}
\hypersetup{urlbordercolor={.5 1 1}}
\hypersetup{linkbordercolor={1 .7 .7}}

\renewcommand{\=}{\mathrel{\phantom{=}}}

\newcommand{\ang}[2]{\langle #1\;#2\rangle}
\newcommand{\bracket}[2]{\langle #1 \vert #2 \rangle}

\allowdisplaybreaks

\begin{document}
\thispagestyle{empty}
%\ifarxiv\vspace*{-20mm}\fi

%\ifarxiv\else\begingroup\raggedleft\footnotesize\ttfamily
%AEI-2010-019\\
\begin{flushright}
HU-EP-12/16
\end{flushright}
%\arxivlink{1002.1733}\par\vspace{15mm}
%\endgroup\fi

\begingroup\centering
{\Large\bfseries\mathversion{bold}
Comparing efficient computation methods for\\ massless QCD tree amplitudes:\\
Closed Analytic Formulae versus
Berends-Giele Recursion
\par}%
\hypersetup{pdfsubject={}}%
\hypersetup{pdfkeywords={}}%
\ifarxiv\vspace{15mm}\else\vspace{15mm}\fi

\begingroup\scshape\large
Simon Badger${}^{1}$, Benedikt Biedermann${}^{2}$,
Lucas Hackl${}^{2,3}$, \\[0.1cm]
Jan Plefka${}^{2}$, Theodor Schuster${}^{2}$ and Peter Uwer${}^{2}$
\endgroup \vspace{5mm}

\begingroup\ifarxiv\small\fi
\textit{${}^{1}$The Niels Bohr International Academy and Discovery Center \\
    The Niels Bohr Institute\\
    Blegdamsvej 17, DK-2100 Copenhagen, Denmark}\\[0.1cm]
\ifarxiv\texttt{simon.badger@nbi.dk}\fi
\endgroup
\vspace{0.5cm}

\begingroup\ifarxiv\small\fi
\textit{${}^{2}$Institut f\"ur Physik, Humboldt-Universit\"at zu Berlin, \\
Newtonstra{\ss}e 15, D-12489 Berlin, Germany}\\[0.1cm]
\ifarxiv\texttt{\{biedermann,lucas.hackl,plefka,uwer,theodor\}@physik.hu-berlin.de\phantom{\ldots}}\fi
\endgroup
\vspace{0.5cm}

\begingroup\ifarxiv\small\fi
\textit{${}^{3}$Perimeter Institute for Theoretical Physics, \\
Waterloo, Ontario N2L 2Y5, Canada  \\[0.2cm]}
\endgroup
\vspace{1cm}

\textbf{Abstract}\vspace{5mm}\par
\begin{minipage}{14.7cm}
  Recent advances in our understanding of tree-level QCD amplitudes in
  the massless limit exploiting an effective (maximal) supersymmetry
  have led to the complete analytic construction of tree-amplitudes
  with up to four external quark-anti-quark pairs.  In this work we
  compare the numerical efficiency of evaluating these closed analytic
  formulae to a numerically efficient implementation of the
  Berends-Giele recursion. We compare calculation times for
  tree-amplitudes with parton numbers ranging from 4 to 25 with no,
  one, two and three external quark lines. We find that the exact
  results are generally faster in the case of MHV and NMHV amplitudes.
  Starting with the NNMHV amplitudes the Berends-Giele
  recursion becomes more efficient. In addition to the runtime
    we also compared the numerical accuracy. The analytic formulae are
    on average more accurate than the off-shell recursion relations
    though both are well suited for complicated phenomenological
    applications. In both cases we observe a reduction in the average
    accuracy when phase space configurations close to singular regions
    are evaluated. We believe that the above findings provide valuable
    information to select the right method for phenomenological
    applications.
\end{minipage}\par
\endgroup
\newpage

%%%%%%%%%%%%%%%%%%%%%%%%%%%%%
%\pagenumbering{arabic}
%\setcounter{page}{1}
%\renewcommand{\thefootnote}{\arabic{footnote}}
%\setcounter{footnote}{0}

\setcounter{tocdepth}{2}
\hrule height 0.75pt
\tableofcontents
\vspace{0.8cm}
\hrule height 0.75pt
\vspace{1cm}

\setcounter{tocdepth}{2}

%%%%%%%%%%%%%%%%%%%%%%%%%%%%%%%%%%%%%%%%%%%%%%%%%%%%%%%%%%%%%%%%%%%%%%%%%%%%%

\section{Introduction}

Numerically fast and accurate computation methods for multi-parton
tree-level amplitudes in QCD are of great importance from many points
of view. They crucially enter the theoretical prediction for cross
sections of multi-jet processes at leading order (LO) in the QCD
coupling $\alpha_{s}$ as they occur at present particle colliders such
as the Large Hadron Collider (LHC).  Here a variety of computer
programs based on the numerical evaluation of Feynman diagrams have
been developed in the past see for example
\cite{Stelzer:1994ta,Krauss:2001iv,Gleisberg:2008fv}. With the LHC data from the
year 2011 jet multiplicities of up to 9 jets in the final state are
probed. With the data of the year 2012, LHC will be able to investigate
jet multiplicities of up to 12 jets. However, for high multiplicities, the conventional Feynman diagram based approach quickly reaches its limit, 
for example 8 jets in the final state would require already the evaluation of more than $10^7$ Feynman diagrams!
Hence more efficient methods are needed. Here important progress has been made in recent years based 
on recursive on-shell methods. Moreover, QCD tree-amplitudes are crucially needed
for the computation of one-loop corrections, when these are
constructed using a numerical implementation of generalized unitarity
(for recent reviews see refs.~\cite{Bern:2007dw,Berger:2009zb,Ellis:2011cr,Ita:2011hi}).
Recently rapid progress has been made in developing and automating the generalized unitarity and
integrand reduction approaches to computation of loop amplitudes \cite{Badger:2010nx,Becker:2010ng,Berger:2008sj,Bevilacqua:2011xh,Cascioli:2011va,Cullen:2011ac,Giele:2008bc,Giele:2009ui,Hirschi:2011pa,Lazopoulos:2008ex}. These techniques have made NLO predicitons
for multi-jet final states at hadron colliders feasible for up to $2\to 5$ processes (for recent results see for example \cite{Bern:2011ep,Berger:2010zx,Ita:2011wn,Bredenstein:2009aj,Denner:2010jp,Frederix:2010ne,Becker:2011vg,Greiner:2012im,Greiner:2011mp,Bevilacqua:2011aa,Bevilacqua:2009zn,Bevilacqua:2010qb,Bozzi:2011en,Campanario:2011ud,Arnold:2011wj,Becker:2012aq,KeithEllis:2009bu,Frederix:2011ig,Melia:2010bm,Melia:2011dw}).
On the formal side a number of new methods have been devised for the
computation of scattering amplitudes in $SU(N_c)$ gauge theories with
particular emphasis on the maximally supersymmetric ($\mathcal{N}=4$)
Yang-Mills theory. Among a host of other developments the Britto,
Cachazo, Feng and Witten (BCFW) recursion
relation~\cite{Britto:2004ap,Britto:2005fq} was developed which uses
only on-shell lower-point amplitudes evaluated at complex momenta to
construct the desired higher-point trees in any gauge theory. The BCFW
recursion was then recast as a super-recursion for super-amplitudes in
the maximally supersymmetric case of ($\mathcal{N}=4$) Yang-Mills
theory \cite{ArkaniHamed:2009dn}. This super-recursion could then be
solved for arbitrary external states by Drummond and Henn
in~\cite{Drummond:2008cr} leading to closed analytic formulae for
all super-amplitudes at tree-level. The projection of the
super-amplitudes on the component field level yielding $\mathcal{N}=4$
super Yang-Mills theory amplitudes with external gluons, gluinos and
scalars may be obtained upon suitable Grassmann integrations. This was
done recently by Dixon, Henn and two of the present authors
\cite{Dixon:2010ik} who went on to show, that all primitive
tree-amplitudes in massless QCD with up to four external quark lines
of arbitrary flavors and an arbitrary number of gluons may be obtained
from associated gluon-gluino trees in $\mathcal{N}=4$ Yang-Mills
theory. This result was then exploited to write down explicit analytic
formulae for all tree-level amplitudes in massless QCD with up to four
quark lines. Moreover, a publicly available  {\sc Mathematica} package {\tt GGT} was provided which generates all analytic tree-level
gluon-gluino amplitudes relevant for QCD.\footnote{A {\sc Mathematica} package for all tree-level amplitudes in N=4 SYM appeared in \cite{Bourjaily:2010wh}.} In its current version GGT directly provides all QCD tree amplitudes with up to six quarks. The obtained analytic
formulae are very compact at the maximally-helicity-violating (MHV) and
next-to-maximally helicity violating (NMHV) levels but do grow
considerably in complexity with growing $k$ for N${}^{k}$MHV
amplitudes.

Hence the ``$\mathcal{N}=4$ SUSY method'' \cite{Dixon:2010ik} for the
evaluation of massless QCD trees based on exact formulae displays a
complementary situation to the conventional Berends-Giele recursive
approach for the efficient numerical evaluation of trees in that its
evaluation time scales mildly with the parton number $n$ but strongly
depends on the number of helicity flips $k$ of the amplitude
considered.  In contrast the Berends-Giele recursion evaluation time
is independent of $k$ but strongly depends on the number of partons
$n$. The purpose of this work is a detailed analysis of the
computation times for these two approaches as well as a test of their
numerical accuracy. The outcome of our analysis may serve as a
guideline on which implementation should be used in order to maximally
speed up the numerical implementation of massless QCD trees in future
numerical calculations. The ability to calculate tree amplitudes numerically in a fast and accurate manner even for high multiplicity
opens up a variety of new applications beyond the current uses in fixed order calculations. Given the recent progress in
the refinement of matching and parton-shower algorithms together with
extended reach of the LHC searches it is very likely
that amplitudes involving ten or even more external partons will enter
phenomenological studies in the future.

A similar analysis as presented in this paper has been
  performed in Refs.~\cite{Dinsdale:2006sq,Duhr:2006iq}. In difference
  to Refs.~\cite{Dinsdale:2006sq,Duhr:2006iq} we focus on the
  comparison of a purely numerical approach with the usage of analytic
  formulae, since this has not been studied in detail
  before. Furthermore we compare the  numerical accuracy of the
  numerical approach with the numerical evaluation of analytic
  formulae. We also note that very recently the usage of Graphics
  Processing Units (GPUs) for the evaluation of tree-amplitudes has been investigated. More details on this
  interesting option can be found for example in 
  Refs.~\cite{Giele:2010ks,Hagiwara:2009cy}.

\section{Description of used methods}

Tree-level gluon amplitudes in non-abelian gauge theories may be
conveniently separated into a sum of terms, each composed of a simple
prefactor containing the color indices, multiplied by a kinematical
factor known as a partial or color-ordered amplitude.  For an
$n$-gluon amplitude one has
\begin{equation}
{\cal A}^{\text{tree}}_{n}(1,2,3,\ldots , n)
= g^{n-2}\sum_{\sigma\in S_{n}/Z_{n}}
\Tr(T^{a_{\sigma(1)}}\ldots T^{a_{\sigma(n)}})\,
A_{n}(\sigma(1)^{h_{\sigma(1)}}\ldots \sigma(n)^{h_{\sigma(n)}} )\, ,
\label{colorordering}
\end{equation}
with the argument $i^{h_{i}}$ of the partial amplitude $A_{n}$
denoting an outgoing gluon of light-like momentum $p_{i}$ and helicity
$h_{i}=\pm 1$, $i\in[1,n]$. The $su(N_c)$ generator matrices $T^{a_i}$
are in the fundamental representation, and are normalized so that
${\rm Tr}(T^a T^b) = \delta^{ab}$. Finally, $g$ is the gauge coupling.

Similarly, the color decomposition of an amplitude with
a single quark-anti-quark pair and $(n-2)$ gluons is
\begin{equation}
{\cal A}^{\text{tree}}_{n}(1_{\bar{q}},2_q,3,\ldots,n)
 \ =\  g^{n-2} \sum_{\sigma\in S_{n-2}}
   (T^{a_{\sigma(3)}}\ldots T^{a_{\sigma(n)}})_{i_2}^{~\ib_1}\
    A_n^\tree(1_{\bar{q}},2_q,\sigma(3),\ldots,\sigma(n))\,.
\label{qqgluecolorordering}
\end{equation}
Amplitudes with two and more quark-anti-quark lines may be obtained
similarly, as explained in ref.~\cite{Mangano:1990by}.
Though the color structure becomes more intricate in these
  cases, all of the
required kinematical terms are constructed from
suitable linear combinations of the color-ordered amplitudes for
$2k$ external gluinos and $(n-2k)$ external gluons.

Color-ordered amplitudes of massless particles are most compactly
expressed in the spinor-helicity formalism. Here all light-like
four-momenta are written as the product of two
component Weyl spinors
\begin{equation}
\pslash^{\alpha\dot\alpha}= \sigma_{\mu}^{\alpha\dot\alpha}\, p^{\mu}
= \lambda^{\alpha}\, \tilde \lambda^{\dot\alpha} ,
\end{equation}
where we take $\sigma^{\mu}=(\mathbf{1},\vec{\sigma})$ with
$\vec{\sigma}$ being the $2\times 2$ Pauli spin matrices. The spinor
indices are raised and lowered with the Levi-Civita tensor,
i.e.~$\lambda_\alpha=\epsilon_{\alpha\beta}\, \lambda^\beta$ and
$\tilde\lambda_{\dot\alpha}=\epsilon_{\dot\alpha\dot\beta}\,
\tilde\lambda^{\dot\beta}$. An explicit representation is 
\begin{equation}
|\lambda\rangle := \lambda_\alpha = \sfrac{\sqrt{p^0+p^3}}{p^1-ip^2}\,
\begin{pmatrix} % or pmatrix or bmatrix or Bmatrix or ...
  p^1-ip^2 \\
  p^0-p^3 \\
\end{pmatrix} \, ,\qquad
|\tilde\lambda] := \tilde\lambda^{\dot\alpha} =
\sfrac{\sqrt{p^0+p^3}}{p^1+ip^2}\,
\begin{pmatrix} % or pmatrix or bmatrix or Bmatrix or ...
  -p^0+p^3 \\ p^1+ip^2 \\
\end{pmatrix} \, .
\label{reducedspinors}
\end{equation}
In our convention all parton momenta are outgoing. The
amplitudes depend on contracted helicity spinors
\begin{equation}
 \vev{i\,
  j} := \vev{\lambda_i\, \lambda_j} := \epsilon_{\alpha\beta}
\, \lambda_i^\alpha\, \lambda_j^\beta \, , \qquad \bev{i\, j}
:= \bev{\lambda_i\, \lambda_j} :=
\epsilon_{\dot\alpha\dot\beta} \,
\tilde\lambda_i^{\dot\alpha}\, \tilde\lambda_j^{\dot\beta}\, ,
\end{equation}
which are Lorentz invariants.

\subsection{Closed analytic formulae}
\label{sec:AnalyticFormulae}

Compact analytic formulae for tree-amplitudes in the spinor helicity
formalism can be classified by the amount of helicity
violation. The simplest class is the maximally-helicity-violating
(MHV) one. Here either two negative helicity gluons, one negative
helicity gluon and one quark-anti-quark pair or two quark-anti-quark
pairs sit at arbitrary positions within positive helicity gluon states
of the color-ordered amplitude. These read for the zero and one
fermion line case \cite{Parke:1986gb,Mangano:1990by}
\begin{align}
A^\text{MHV}_n(a^-,b^-) &= \delta^{(4)}(p)\, \frac{\ang{a}{b}^4}{\ang{1}{2}\, \ang{2}{3}\ldots\ang{n}{1}} \\
A^\text{MHV}_n(a^-, b_q,c_{\bar{q}})
&= \delta^{(4)}(p)\frac{\ang{a}{c}^3\ang{a}{b}}
{\ang{1}{2}\dots\ang{n}{1}} \,, \\
A^\text{MHV}_n(a^-,b_{\bar{q}},c_q)
&=-\delta^{(4)}(p)\frac{\ang{a}{b}^3\ang{a}{c}}
{\ang{1}{2}\dots\ang{n}{1}}\, ,
\end{align}
where $a,b,c\in [1,n]$ and $a^-$ denotes a negative-helicity gluon at
position $a$, while fermions of opposite helicity with flavors $A$, $B$ at positions $b$, $c$ 
are denoted by $b^A_q$ ($+\frac12$) and $c^B_{\bar{q}}$ ($-\frac12$), and
$p=\sum_{i=1}^n p_i$. Flavor becomes important for the two fermion
line case where one has three distinct representatives depending on
helicity and flavor distributions \cite{Dixon:2010ik}
\begin{align}
A^\text{MHV}_n(a_q^1, b_q^2,c_{\bar{q}}^2,d_{\bar{q}}^1)
&= -\delta^{(4)}(p)\frac{\ang{c}{d}^3\ang{a}{b}}
{\ang{1}{2}\dots\ang{n}{1}}=A^\text{MHV}_n(a_q, b_q,c_{\bar{q}},d_{\bar{q}}) \,, \\
A^\text{MHV}_n(a^1_q, b_{\bar{q}}^1,c_q^2,d_{\bar{q}}^2)
&=\delta^{(4)}(p)\frac{\ang{b}{d}^2\ang{d}{a}\ang{c}{b}}
{\ang{1}{2}\dots\ang{n}{1}}\, , \\
A^\text{MHV}_n(a_{q}, b_{\bar{q}},c_q,d_{\bar{q}})
&=\delta^{(4)}(p)\frac{\ang{b}{d}^3\ang{a}{c}}
{\ang{1}{2}\dots\ang{n}{1}}\, ,
\end{align}
where the flavor index has been omitted in the single flavor cases.

The complexity of the closed formulae grows at the NMHV level
comprising color-ordered amplitudes with $k$ negative helicity gluons
and $3-k$ quark-anti-quark pairs embedded in a sea of $(n+k-6)$
positive helicity gluon states. In order to express the formulae in a
compact way one needs to introduce the region momenta
$x_{ij}^{\alpha\dot\alpha}$ via
\begin{equation}
\label{12}
x_{ij}^{\alpha \dot{\alpha}}
\ :=\ ( \pslash_{i}+ \pslash_{i+1} + \cdots
 + \pslash_{j-1} )^{\alpha \dot{\alpha}}
= \sum_{k=i}^{j-1}\, \lambda_{k}^{\alpha} \tilde{\lambda}_{k}^{\dot\alpha}
 \, , \qquad
\, \qquad i<j \, ,
\end{equation}
$x_{ii} = 0$, and $x_{ij} = -x_{ji}$ for $i>j$.
All N${}^k$MHV tree-level amplitudes can be expressed in terms of the
quantities $\langle n a_{1} a_{2} \ldots a_{k} |a\rangle$ defined by
\begin{equation}
\langle n a_{1} a_{2} \ldots a_{k} |a\rangle :=
\langle n| x_{na_{1}} x_{a_{1}a_{2}}\ldots x_{a_{k-1}a_{k}}|a\rangle \, ,
\label{defvev}
\end{equation}
and the spinor products $\ang{i}{j}$. The pure gluon NMHV amplitude
with negative helicity gluons sitting at positions $a,b$ and $n$ takes
the form \cite{Dixon:2010ik}
\begin{multline}
A_n^{\mathrm{NMHV}}(a^-,b^-,n^-)
=\frac{\delta^{(4)}(p)}{\ang{1}{2}\dots \ang{n}{1}}\times \\
\begin{aligned}
 &  \Bigg[  \sum_{a < s \leq b < t\leq n-1}\tilde{R}_{n;st}
   \Bigl(   \ang{n}{a} \vev{nts|b}  \Bigr)^{4}
        +    \sum_{a < s <   t \leq b}\tilde{R}_{n;st}
   \Bigl( \ang{b}{n} \ang{n}{a} x_{st}^2 \Bigr)^{4}  \\
     &+\sum_{2\leq  s \leq a < b <   t  \leq n-1}\tilde{R}_{n;st}
   \Bigl( \ang{ b}{a} \vev{n ts|n} \Bigr)^{4}
         + \sum_{2\leq  s \leq a   <   t \leq b}\tilde{R}_{n;st}
   \Bigl(  \ang{n}{b} \vev{nst|a} \Bigr)^{4}
        \Bigg]  \,,
\end{aligned}
\end{multline}
where we have introduced the $\tilde R$-invariant
\begin{equation}
\tilde R_{n; st}:= \frac{1}{x_{st}^{2}}\,
\frac{\vev{s(s-1)}}{\langle n t s |s\rangle\,
\langle n t s |s-1\rangle}
\frac{\vev{t(t-1)}}{\langle n s t |t\rangle \,
\langle n s t |t-1\rangle} \, .
\label{Rtilde1}
\end{equation}
with $\tilde R_{n; st} := 0$ for $t=s+1$ or $s=t+1$.  Let us also
state NMHV amplitude with one quark-anti-quark pair and two negative-helicity gluons.
Here there are two distinct configurations (we take $a<b<c$ below)
\cite{Dixon:2010ik}
\begin{align}
A_n^{\mathrm{NMHV}}(a_q,b^-,c_{\bar{q}},n^-)&=\frac{\delta^{(4)}(p)}
{\ang{1}{2}\dots\ang{n}{1}}\biggl[
{}-\ang{a}{b}\ang{b}{c}^3\sum_{1<s\leq a,b,c<t< n}
\bracket{nts}{n}^4\tilde{R}_{n,st} \notag\\
&\=\phantom{\frac{\delta^{(4)}(p)}
{\ang{1}{2}\dots\ang{n}{1}}\biggl[}
-\ang{b}{c}^3\ang{a}{n}\sum_{a<s\leq b,c<t<n}
\bracket{nts}{b}\bracket{nts}{n}^3\tilde{R}_{n,st}\notag\\
&\=\phantom{\frac{\delta^{(4)}(p)}
{\ang{1}{2}\dots\ang{n}{1}}\biggl[}
-\ang{c}{n}^3\ang{a}{n}\sum_{a<s\leq b<t\leq c}
\bracket{nst}{b}^3\bracket{nts}{b}\tilde{R}_{n,st}\notag\\
&\=\phantom{\frac{\delta^{(4)}(p)}
{\ang{1}{2}\dots\ang{n}{1}}\biggl[}
-\ang{c}{n}^3\ang{a}{b}\sum_{1<s\leq a,b<t\leq c}
\bracket{nst}{b}^3\bracket{nts}{n}\tilde{R}_{n,st}\biggr]\,,
\end{align}
\begin{align}
A_n^{\mathrm{NMHV}}(a_q,b_{\bar{q}},c^-,n^-)&=\frac{\delta^{(4)}(p)}
{\ang{1}{2}\dots\ang{n}{1}}\biggl[{}
+\ang{a}{c}\ang{b}{c}^3\sum_{1<s\leq a,b,c<t< n}
\bracket{nts}{n}^4\tilde{R}_{n,st}\notag\\
&\=\phantom{\frac{\delta^{(4)}(p)}
{\ang{1}{2}\dots\ang{n}{1}}\biggl[}
+\ang{a}{n}\ang{b}{n}^3\sum_{b<s\leq c<t<n}
\bracket{nts}{c}^4\tilde{R}_{n,st}\notag\\
&\=\phantom{\frac{\delta^{(4)}(p)}
{\ang{1}{2}\dots\ang{n}{1}}\biggl[}
+\ang{n}{c}^4\ang{a}{n}\ang{b}{n}^3\sum_{b<s<t\leq c }
(x_{st}^2)^4\tilde{R}_{n,st}\notag\\
&\=\phantom{\frac{\delta^{(4)}(p)}
{\ang{1}{2}\dots\ang{n}{1}}\biggl[}
+\ang{c}{n}^4\sum_{1<s\leq a,b<t\leq c}
\bracket{nst}{b}^3\bracket{nst}{a}\tilde{R}_{n,st}\notag\\
&\=\phantom{\frac{\delta^{(4)}(p)}
{\ang{1}{2}\dots\ang{n}{1}}\biggl[}
+\ang{b}{c}^3\ang{a}{n}\sum_{a<s\leq b,c<t<n}
\bracket{nts}{c}\bracket{nts}{n}^3\tilde{R}_{n,st}\notag\\
&\=\phantom{\frac{\delta^{(4)}(p)}
{\ang{1}{2}\dots\ang{n}{1}}\biggl[}
+\ang{c}{n}^4\ang{a}{n}\sum_{a<s\leq b<t\leq c}
x_{st}^2\bracket{nst}{b}^3\tilde{R}_{n,st}\biggr]\,.
\end{align}
The two and three fermion line amplitudes are more involved and may be found in the appendix
B of [9]. These formulae have also been implemented in the publicly available {\sc Mathematica}
package {\tt GGT}\footnote{Included in this submission and
  at {\tt http://qft.physik.hu-berlin.de} }. As already stated in the introduction, the current version of {\tt GGT} also provides all QCD amplitudes with up to six quarks.

An estimate for the evaluation time of these closed formulae for the
amplitudes is the number of terms the expression has. From this one
estimates the evaluation time of the MHV amplitudes to be of order $n$ since 
the number of spinor products to be evaluated is approximately $n$. 
The number of terms in the formulae for the NMHV amplitudes grows as $n^2$ for large parton
numbers $n$. Excluding the MHV prefactor the complexity of each of the terms is independent of the 
parton number, hence the asymptotic scaling in evaluation time
is $n^2$.  This is competitive with the Berends-Giele recursion
method which grows independent on the helicity distributions of the
partons as $n^4$, discussed in the next subsection. The NNMHV
formulae of \cite{Dixon:2010ik} display a growth in the number of
terms as $n^4$. Due to the same arguments as in the NMHV case we thus expect
a similar performance of the NNMHV formulae as the Berends-Giele
recursion and a detailed comparison is needed to see which method
wins.  Going beyond the NNMHV level with the analytic formulae of \cite{Dixon:2010ik} for the
amplitudes appears to be disfavored as in general the number of terms in an N${}^k$MHV formula grows as $n^{2k}$
for large parton numbers.

The closed analytic formulae of \cite{Dixon:2010ik} for the MHV,
NMHV and NNMHV with zero to three quark-anti-quark lines have been
directly implemented in a {\tt C++} program {\tt cGGT.cpp} which can be provided upon request. {\tt
  cGGT.cpp} contains the straightforwardly hard-coded analytic
formulae and a natural amount of caching is performed in order to
speed up the numerical evaluation of the amplitudes for a given
phase-space point. As such all the region momenta are evaluated and
stored during initialization, similarly all
spinor brackets are evaluated with the reduced spinors in
\Eqn{reducedspinors} without the square root dependent pre-factor,
which is only evaluated at the very end, as typically even powers of
the pre-factor arise.

\subsection{Berends-Giele recursion}
In this subsection we briefly comment on a purely numerical
implementation of leading-order scattering amplitudes in massless QCD.
Since an extensive literature exists on the subject we limit ourselves
to the basic ingredients. More details can be found in
Ref.~\cite{Berends:1987me}. In difference to on-shell recurrence
relations developed more recently \cite{Britto:2004ap,Britto:2005fq},
the Berends-Giele recursion uses off-shell currents as basic building
blocks. In pure gauge theory the off-shell currents
$J_\mu(1^{h_1},2^{h_2}\ldots,n^{h_n})$ correspond to the amplitudes
for the production of $n$ gluons with helicities $h_i$ and one
off-shell gluon with the corresponding polarization vector stripped
off. The on-shell scattering amplitude is
thus obtained by taking the on-shell limit and contracting with the
polarization vector of the additional gluon. As mentioned in the
previous section it is convenient to split the scattering amplitude
into a color part and the remaining Lorentz structure. In practice
this can be done for example by using color-ordered Feynman rules (for
details we refer to Ref.~\cite{Dixon:1996wi}). The full amplitude will
in general contain different color structures.  However since not all
of these structures are independent it is usually sufficient to calculate only
a few of them and reconstruct the remaining ones by permuting the
external gluons.  The key observation leading to the Berends-Giele
recurrence relation is the fact that any off-shell current can be
written as a sum of simpler off-shell currents connected via the
appropriate three- ($V_{3g}$) and four-gluon ($V_{4g}$) vertices:
\begin{align}
  J_\mu(1,...,n) &= \frac{-i}{P_{i,n}^2}
  \left[
    \sum_{i=1}^{n-1} V_{3g}^{\mu\nu\rho}(P_{1,i},P_{i+1,n})
    \,J_\nu(1,...,i)\,J_\rho(i+1,...,n)\right. \nonumber\\
  &\= \phantom{\frac{-i}{P_{i,n}^2}[}+\left.\sum_{j=i+1}^{n-1} \sum_{i=1}^{n-2}
    V_{4g}^{\mu\nu\rho\sigma}\,J_\nu(1,...,i)J_\rho(i+1,...,j)
    \,J_\sigma(j+1,...,n)
  \right]\label{eq:BG-master}
\end{align}
where we have suppressed the helicity index and the gauge coupling is
set to one. In addition the definition
\begin{equation}
  P_{i,j} = \sum_{k=i}^j p_k\qquad\text{for}\quad j\ge i
\end{equation}
is used, we note $P_{i,j}=x_{i,j+1}$ from (\ref{12}).
The color-ordered vertices are given by
\begin{align}
  V_{3g}^{\mu\nu\rho}(P_{1},P_{2})& =
  \frac{i}{\sqrt{2}}\left( g^{\nu\rho}(P_1-P_2)^\mu+2g^{\rho\mu}
    P_2^\nu-2g^{\mu\nu}P_1^\rho\right),\nonumber\\
  V_{4g}^{\mu\nu\rho\sigma} & =
  \frac{i}{2}(2g^{\mu\rho}g^{\nu\sigma}-g^{\mu\nu}g^{\rho\sigma}
  -g^{\mu\sigma}g^{\nu\rho}).
\end{align}
Since the right hand side of \Eqn{eq:BG-master} is formally
simpler---only off-shell currents with a lower number of gluons are
involved---\Eqn{eq:BG-master} can be used to calculate off-shell
currents recursively. The end-point of the recursion is given by
\begin{equation}
  J_\mu(i^{\,h_i}) = \left(\varepsilon_\mu^{(h_i)}(p_i)\right)^{\ast}
\end{equation}
where $\varepsilon_\mu^{(h_i)}(p_i)$ denotes the polarization vector
of a gluon with momentum $p_i$ and polarization $h_i$. We take all the
partons as outgoing, that is the on-shell limit of the scattering
amplitudes correspond to the transition $0\to g(1^{h_1})\cdots
g(n^{h_n})g((n+1)^{h_{n+1}})$. Scattering amplitudes for physical
processes are obtained as usual by crossing. Using the explicit form
of the three- and four-point vertices as given above, the
implementation of \Eqn{eq:BG-master} in a computer program is straight
forward. As can be seen from \Eqn{eq:BG-master} the same sub-current
may appear at different depths of the recursion. To speed up the
numerical evaluation it is thus important to cache the sub-currents and
evaluate them only once. We note that the possibility to reuse
sub-currents during the calculation is a major advantage of off-shell
recurrence relations compared to on-shell methods. Since recursive
implementations tend to be sub-optimal to get high computing
performance \Eqn{eq:BG-master} is implemented as a bottom--up
approach. The program uses the one-point currents specified by
the user in terms of particular polarization states together with the
respective momenta to calculate the two-gluon off-shell currents
$J_\mu(i^{h_i},(i+1)^{h_{i+1}})$. The two-point currents together with
the one-point currents are then used in the subsequent step to calculate the
three-point currents. This procedure is repeated until the current of
maximal length is obtained. Owing to our restriction to specific color
structures only a fixed cyclic ordering needs to be considered. One
can show that if sub-currents are cached the computational effort for
the evaluation of an $n$-point current scales as $n^4$. (Without cache
the scaling would be $4^n$.)  We will come back to this point when we
discuss the numerical performance. As a technical detail we remark
that in the implementation presented here \cite{Badger:2010nx} no
specific bases for the polarization vectors has been used. In
particular no helicity methods have been applied. Since in (almost) all
phenomenological applications the gluon polarization is not observed,
only matrix elements squared summed over all polarization states will
occur. As a consequence an arbitrary bases can be used as long as the
sum over all possible polarization states is complete.  Using real
polarization vectors could thus yield a significant speed up since the
entire calculation can be done using real numbers instead of complex
arithmetic.

The extension to include also quarks---massive as well as massless
ones---is straight forward. The main difference, namely that some
sub-currents do not exist since there is no direct coupling between
quarks, is merely a matter of bookkeeping. We stress that the quark
currents calculated in the way described above in general do not
correspond to partial amplitudes. However partial
amplitudes can be constructed from the aforementioned currents. The
reconstruction of the full matrix elements---not subject of this
article---has been checked for a variety of different processes
\cite{Badger:2012dd}.

\section{Performance and Numerical Accuracy}

The scattering amplitudes described in the previous section find their
application in leading-order phenomenology at hadron colliders.
However, this is not the only application. With the development of
unitarity inspired techniques, leading-order
amplitudes represent an important input to the evaluation of one-loop
amplitudes. In both cases the amplitudes need to be evaluated
for millions of phase space points. The required computation time is
thus an important factor in choosing the optimal approach. We compare
the evaluation time in detail in \Secref{subsec:speed}.

In particular when using leading-order amplitudes in the evaluation of
one-loop amplitudes, not only the speed but also the numerical accuracy
matters. In the unitarity method the one-loop amplitude is
reconstructed from a large number of different cuts requiring the
evaluation of the corresponding tree amplitudes. It is thus important
to assure a good accuracy of the individual contributions.  Even in
the case that analytic formulae are available one should keep in mind
that when it comes to the numerical evaluation usually only a finite
floating point precision is employed --- unless special libraries to
allow for extended precision are used. As a consequence, numerical
cancellations between individual contributions may result in a loss of
accuracy of the final result. Since a detailed understanding of the
numerical uncertainties is also important when results from different
methods are compared we investigate the numerical uncertainties of the
two approaches discussed in the previous Section in
\Secref{subsec:accuracy}.

\subsection{Evaluation Time}
\label{subsec:speed}
\begin{figure}[ht]
   \centering
   \includegraphics[width=14cm]{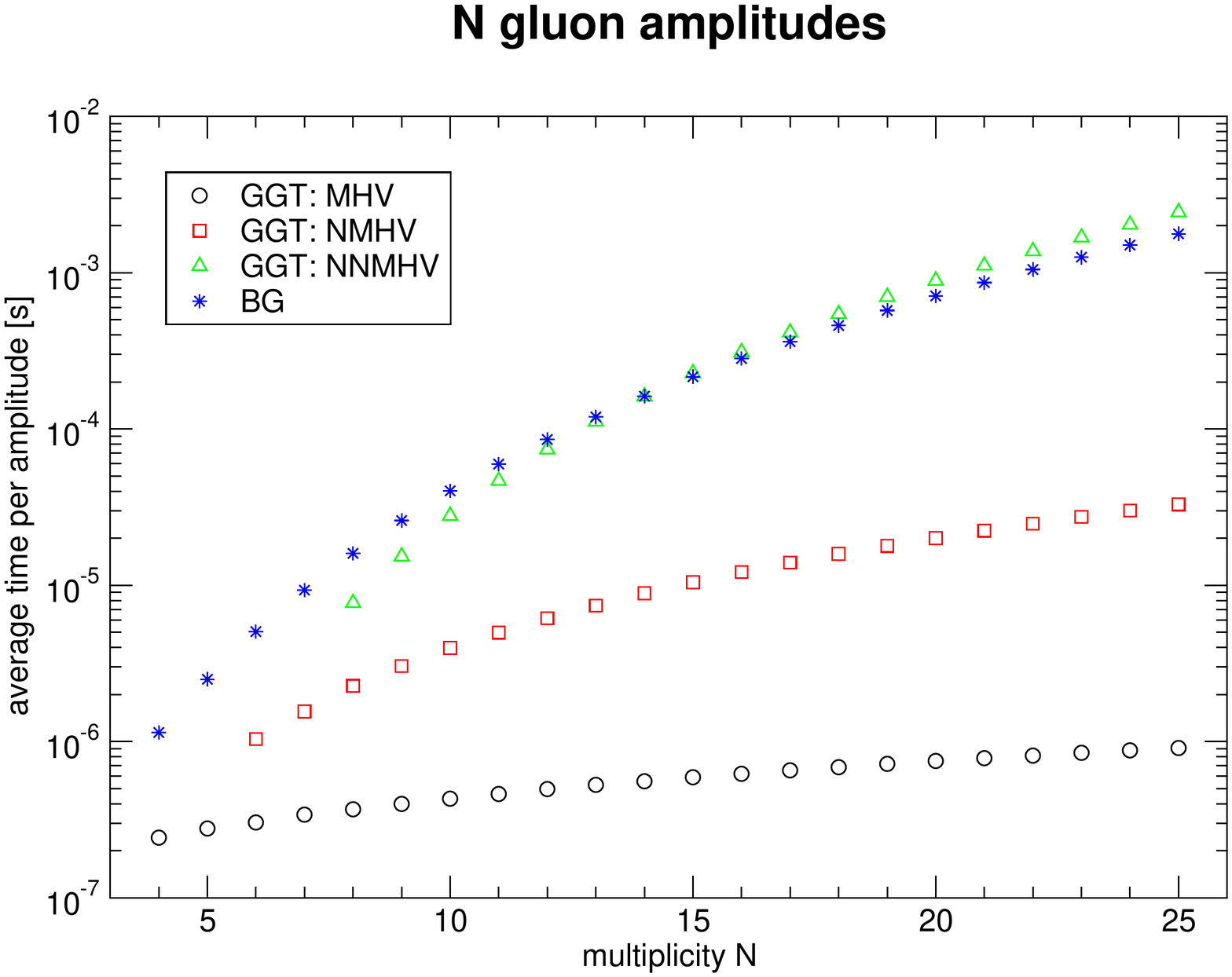}%[angle=-90,width=14cm]{gluons}
   \caption{Average time required per phase space point for the
     evaluation of pure gluon amplitudes as function of the parton
     multiplicity.}
\label{fig:PureGluon}
\end{figure}
Before discussing the results in detail we briefly describe how the runtime is
analyzed. To investigate the performance we used a computer with 16 GByte main memory 
and Intel(R) Core(TM) i5 3.33GHz cpu running under Debian 6.04.
To reduce context switches as much as possible we payed attention to the fact
that the computer was used exclusively for the performance measurements. Furthermore we used the POSIX function {\tt
getrusage} for the measurement of the used cpu time, which is to some great extent context independent. The function returns the time spent in user mode split into seconds
and micro seconds. It is not documented whether the underlying clock provides a
real time accuracy at the level of micro seconds. One can assume however that a
precision at the level of milli seconds should be feasible which is sufficient
for our purpose using the procedure described in the following.

The key observation is that both the evaluation time of the analytical formulae and of the Berends-Giele recusion depend on the positions of the fermions. In the case of the analytical formulae we additionally have a dependence on the position of the negative helicity gluons. Hence, we chose to average over all configurations to which the analytical formulae directly apply without exploiting the cyclic symmetry of the amplitudes, e.g. all configurations with a negative helicity gluon at position $n$ for amplitudes with at least one gluon of negative helicity.  
To obtain reproduceable results and to reduce the computational effort to a
minimum we took the following approach: Per measurement a minimum cpu time of
at least one second is required to obtain reliable results. Using empirical
knowledge together with the known scaling of the runtime as a function of the
multiplicity we estimated the number of phase space evaluations for each
sub-process/multiplicity.  We then generated one phase space point and
evaluated all matrix elements corresponding to our desired average the required number of times. While
in the determination of the accuracy it is important that on-shell condition
and four momentum conservation are respected as precise as possible, the
runtime measurement is insensitive to the ``quality'' of the phase space
point---as long as no floating exceptions are encountered. (Floating point
exceptions would lead to exception handling and the execution of different
code.)   In \figref{fig:PureGluon}
the cpu time per phase space point for pure gluon amplitudes is shown. We
compare analytic formulae for three different helicity configurations (MHV,
NMHV, NNMHV) with the purely numerical approach using the Berends-Giele recursion.
Since in the implementation of the Berends-Giele recursion no helicity methods
are used the runtime is the same for different helicity configurations. Fitting the last five data points with $f(n)=A \,n^B$, where $n$ is the gluon multiplicity, we obtain $B\approx4.12$ which is allready quite close to the predicted asymptotic ${\cal O}(n^4)$ behavior. 
We stress that this is a property of the algorithm and cannot be changed by a different implementation. 
The implementation can only affect the normalization factor in front of the $n^4$
behavior. Let us now compare with the runtime required for the evaluation of
the analytic formulae. In case of the MHV amplitude the evaluation is more than
three orders of magnitude faster for 25 gluons---as one would have expected
given the compactness of the analytic results. We have checked that the timings shown for the MHV amplitudes perfectly agree with the predicted $n^1$ scaling. We emphasize that no time
consuming square roots (contained in the spinor products) have to be evaluated in all gluon amplitudes since each spinor appears an even number of times. This is no longer true for amplitudes involving fermions as their associated spinors appear an odd number of times. Hence, for each fermion one square root is required. 
The predicted large $n$ behavior of $n^2$  for the NMHV
amplitudes is in good agreement with the $n^{2.2}$ fit
from the last five data points in \figref{fig:PureGluon}. This is
still much better than the $n^4$ of the Berends-Giele approach. As a
consequence for large multiplicities the analytic results are almost two orders
of magnitude faster than Berends-Giele.  The situation changes when it comes to
the NNMHV amplitudes. From the number of terms in the analytic expression we
expect an asymptotic behavior of the form $n^4$ leading to a similar rise 
of the runtime as a function of the gluon multiplicity as
observed in the Berends-Giele case. However, fitting the last five data points reveals that, with a scaling of $n^{4.5}$ the analytic formulae are still farther away from the asymptotic behavior than Berends-Giele. Consequently for 15 gluons and more the
Berends-Giele recursion starts to become more efficient.  As mentioned already,
the asymptotic behavior is a property of the underlying algorithm and cannot be
changed by a `more clever' implementation. 

Let us add at that point a remark concerning the absolute timings: For low
multiplicities the evaluation time is of the order of micro seconds while for
$n=25$ order milli seconds are required. For practical applications one should
keep in mind, that the timings are for specific color and spin configurations.
While for low multiplicities the number of color and spin configurations is
still small (i.e. for $n\le5$ only MHV amplitudes exist) one can expect that
color and spin summed squared amplitudes can be evaluated in less than one
milli second per phase space point.  However for large multiplicities the
number of color and spin configurations grows rapidly. A naive sum over color
and spin would thus give an additional factor which would render a brute force
evaluation impossible given today's computing resources. In such cases refined
methods like for example Monte Carlo sums over spins and colors would be
required.  

\begin{figure}[t]
   \centering
   \includegraphics[width=14cm]{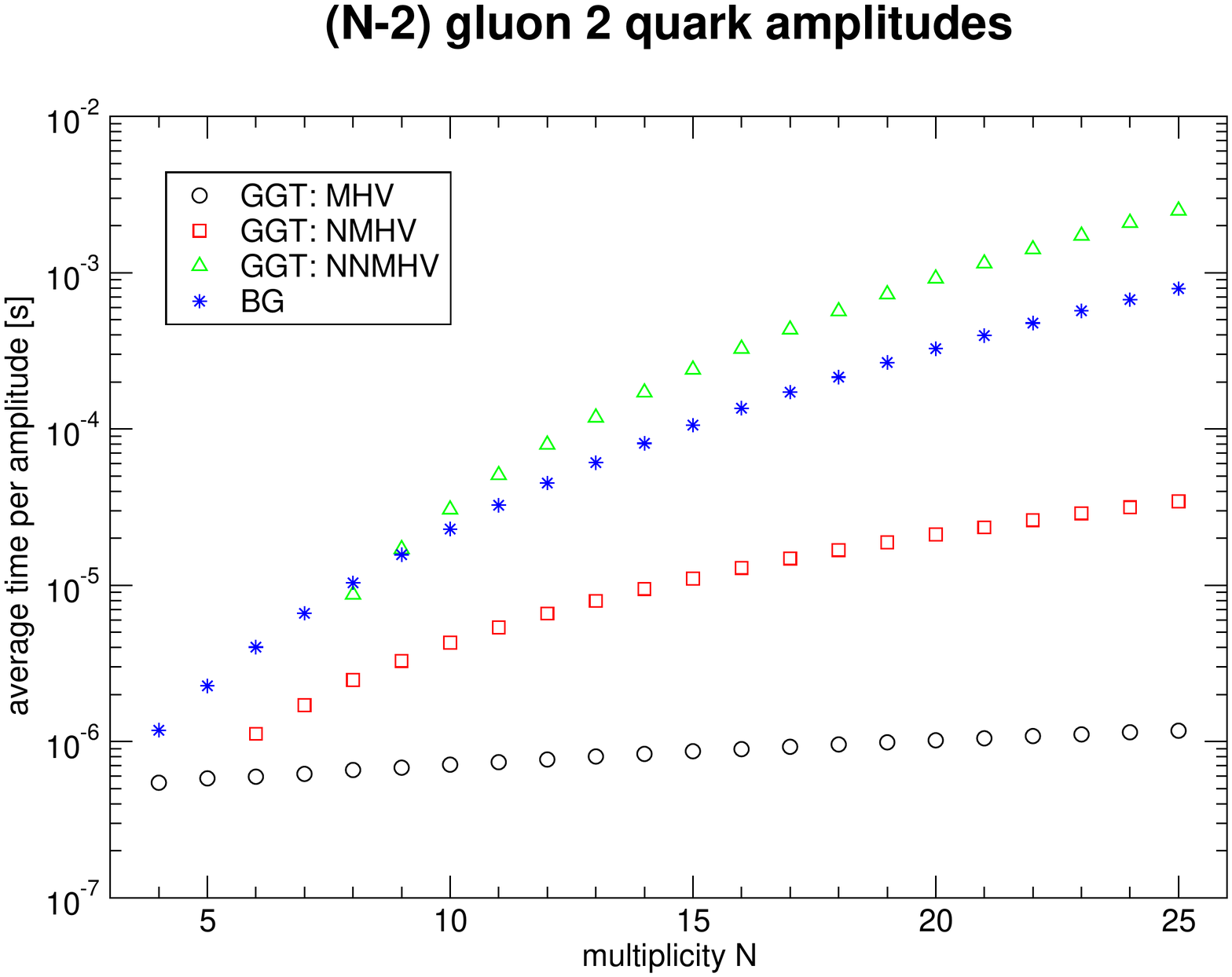}%[angle=-90,width=14cm]{2quark}
     \caption{Evaluation time per phase space point for amplitudes
       with a quark--anti-quark pair and $N-2$ gluons.}
   \label{fig:2Q+Gluons}
\end{figure}

\begin{figure}[ht]
   \centering
   \includegraphics[width=14cm]{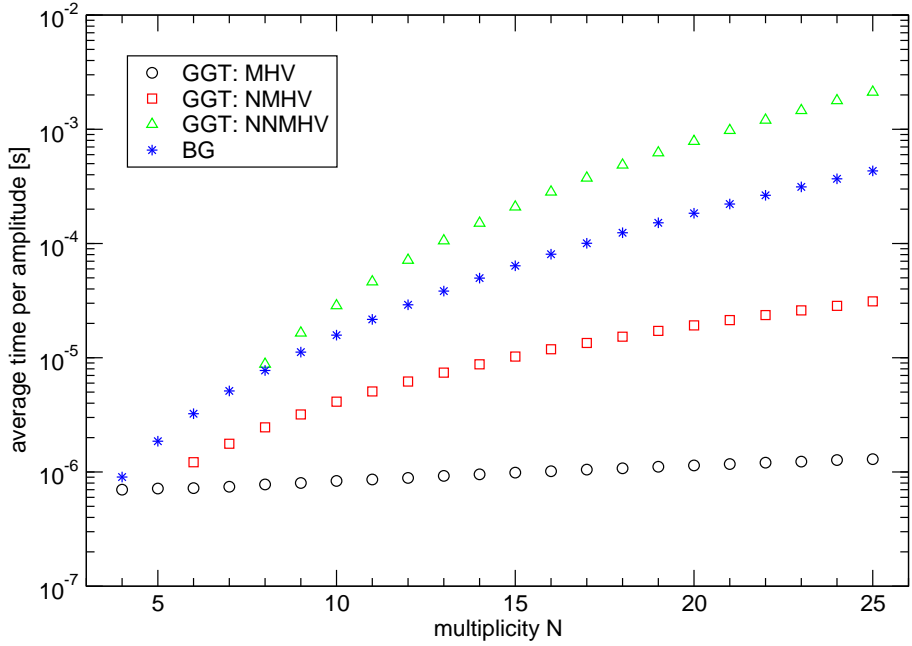}%[angle=-90,width=14cm]{4quark}
     \caption{Evaluation time per phase space point for amplitudes
       with two quark--anti-quark pairs (different flavors) and $N-4$ gluons.}
   \label{fig:4Q+Gluons}
\end{figure}

\begin{figure}[ht]
   \centering
   \includegraphics[width=14cm]{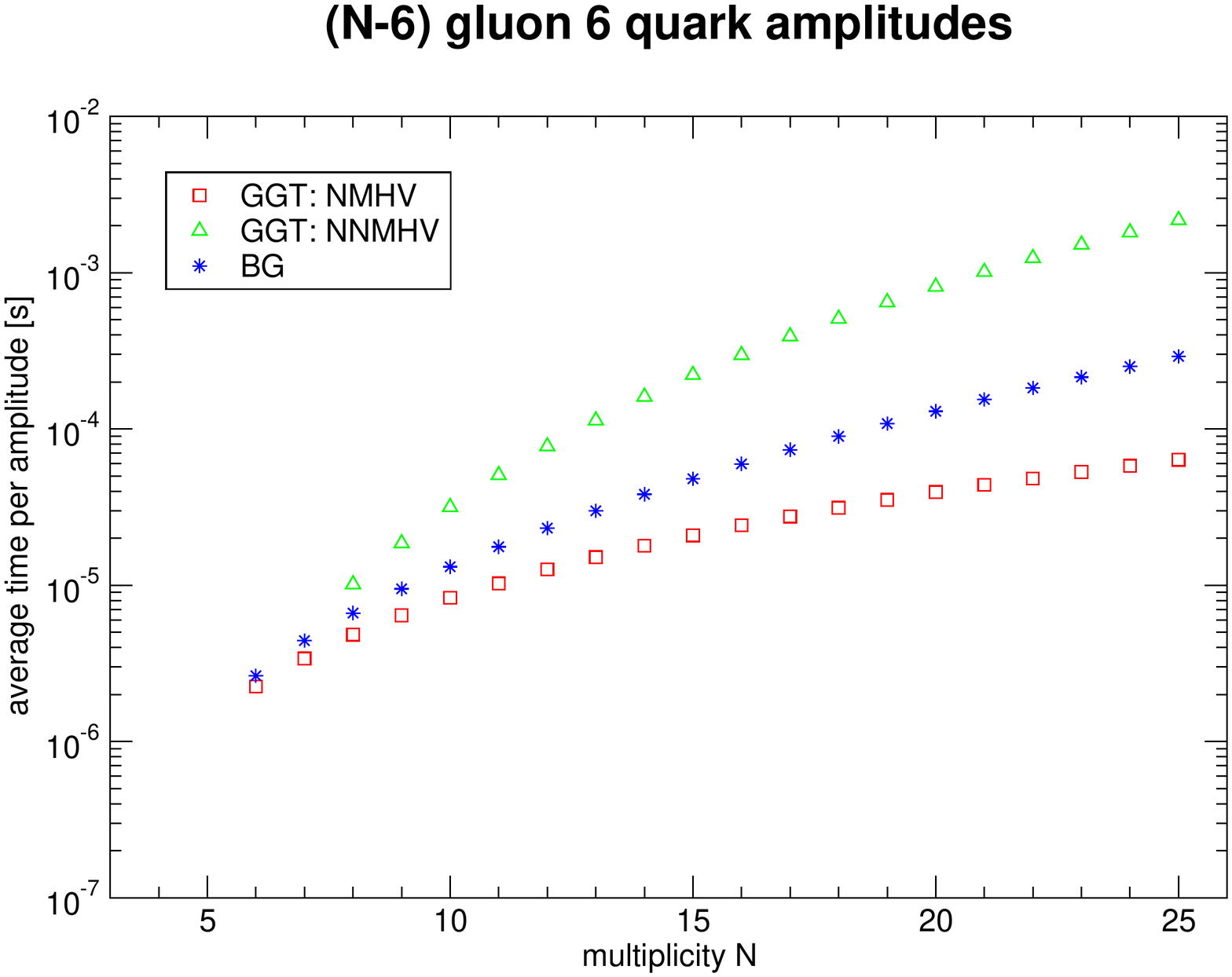}%[angle=-90,width=14cm]{6quark}
   \caption{Evaluation time per phase space point for amplitudes
       with three quark--anti-quark pairs (different flavors)
       and $N-6$ gluons.}
\label{fig:6Q+Gluons}
\end{figure}

In \figref{fig:2Q+Gluons}, \figref{fig:4Q+Gluons} and \figref{fig:6Q+Gluons} we show the results of a similar analysis,
now for amplitudes involving up to three quark--anti-quark pairs.
Again the Berends-Giele recursion method is presented only for a fixed number of negative helicity gluons since our implementation is independent of the gluon helicities. However, to take into account that the runtime depends on
the position of the quarks in the primitive amplitude we took the same configuration average
as for the corresponding analytic formula of smallest MHV degree.
Overall we observe a picture similar to the pure gluon case: for MHV and
NMHV amplitudes the analytic results are much faster than the
evaluation based on the Berends-Giele recursion. Comparing the performance of the Berends-Giele recursion for 0, 2, 4, 6 quarks we find a decreasing dependence on the parton multiplicity. This is simply due to the fact that for a fixed multiplicity the number of currents which have to be evaluated decreases if more fermions are involved. Since the $n^4$ asymptotic of the recursion is due to the four gluon vertex, we expect that the asymptotic scaling will be approached from below. Indeed, for two, four, six quarks we get $n^{3.96}$, $n^{3.83}$, $n^{3.64}$ from the last five data points compared to $n^{3.77}$, $n^{3.43}$, $n^{3.19}$ for up to $n=15$ partons. The timings of the analytical formulae show only a small dependence on the number of quarks. As a consequence the Berends-Giele
recursion is more efficient for the NNMHV amplitudes involving quarks.
In case of all MHV amplitudes it is
remarkable that the analytic formulae for MHV amplitudes show a very
weak dependence on the parton multiplicity. The evaluation of an MHV amplitude for $n=25$ takes only $6\times10^{-7}s$ longer than the evaluation of the four point amplitude. 
This is
easily understood from the structure of formulae since increasing the multiplicity by one results only in the additional evaluation of two reduced spinor products and one squared spinor normalization factor.

\subsection{Numerical Accuracy}
\label{subsec:accuracy}
\begin{figure}[htb]
  \begin{center}
    \leavevmode
    \includegraphics{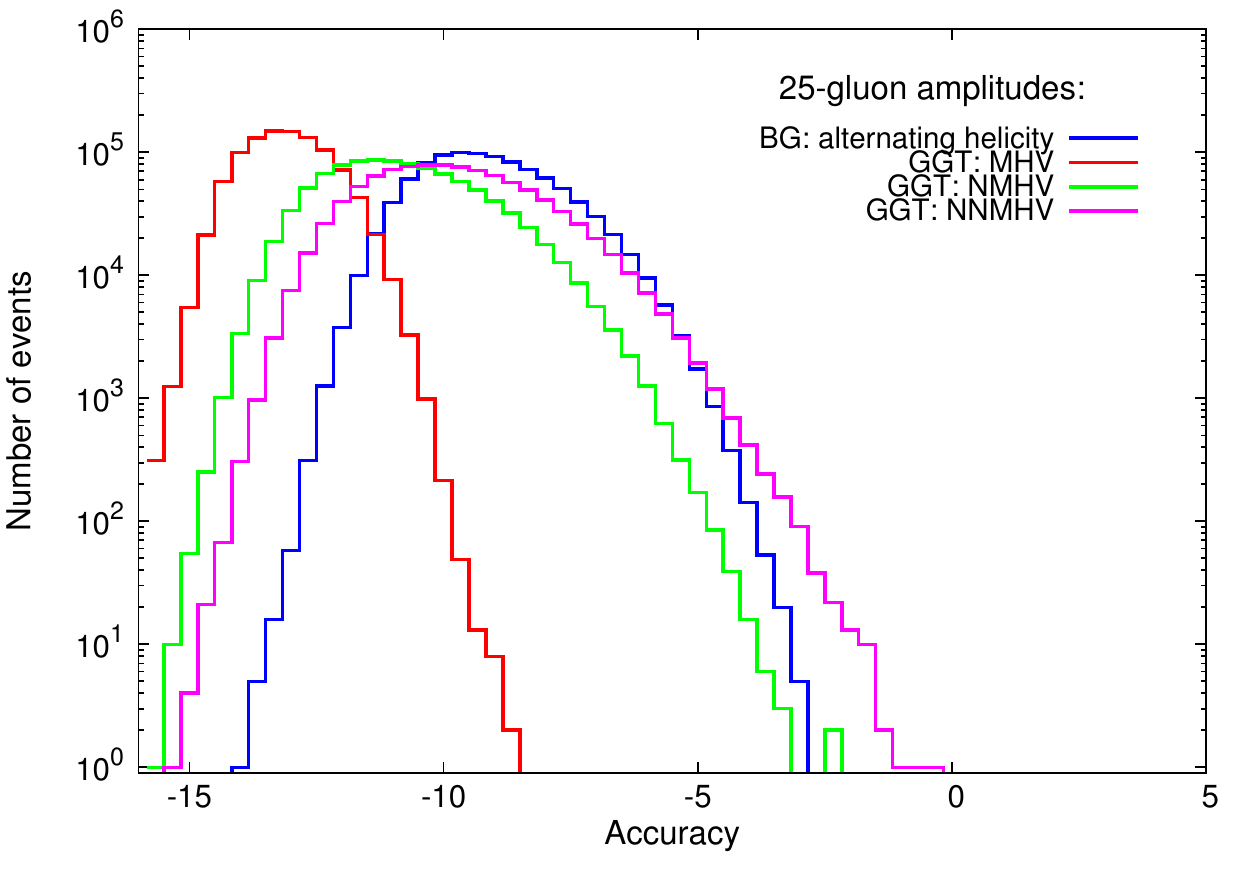}
    \caption{Accuracy $\delta$ for 25 gluon amplitude for purely numerical
    evaluation based on the Berends-Giele recursion (BG) and for
    analytic formulae (GGT) as described in Section
    \ref{sec:AnalyticFormulae}. Phase-space generation by sequential splitting.}
    \label{fig:AccPlotSample}
  \end{center}
\end{figure}

\begin{figure}[htb]
  \begin{center}
    \leavevmode
    \includegraphics[width=14cm]{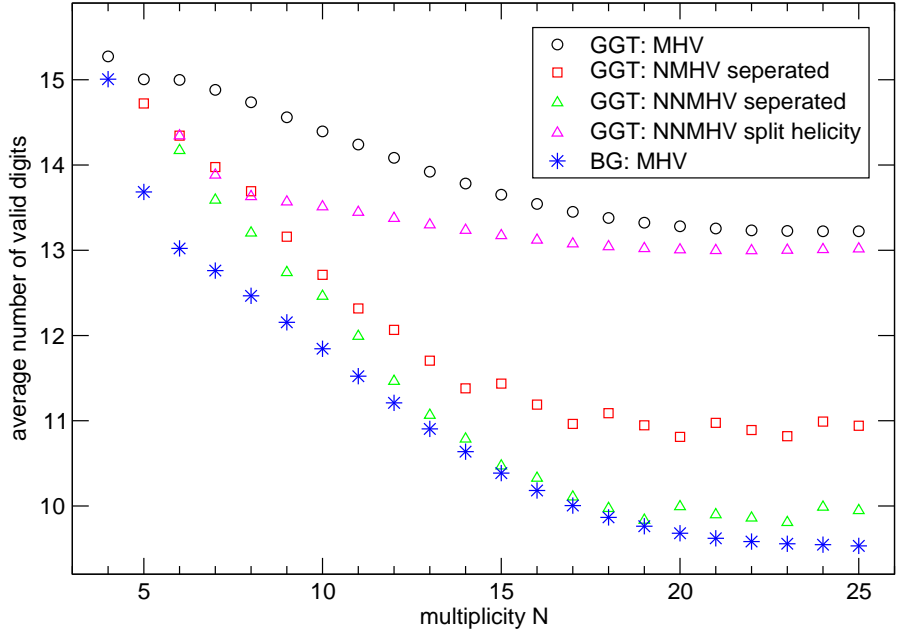}%[angle=-90,width=14cm]{accuracy_gluons}
    \caption{\label{fig:AvgAccGluons}
      Average accuracy $|\delta|$ for pure gluon amplitudes as a function of the
      gluon multiplicity (GGT analytic formulae, BG Berends-Giele).
       Phase-space generation by sequential splitting. }
  \end{center}
\end{figure}

\begin{figure}[htb]
  \begin{center}
    \leavevmode
    \includegraphics{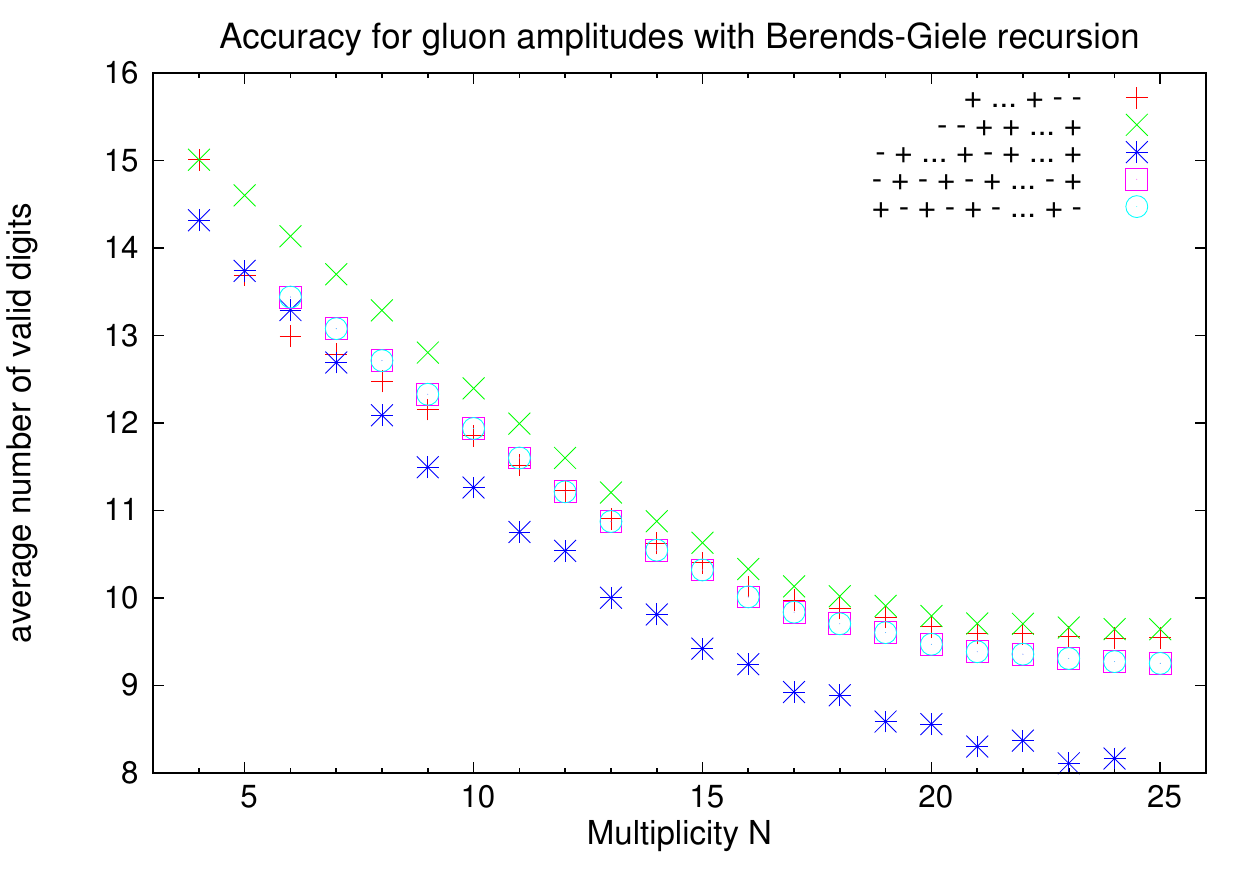}
    \caption{\label{fig:BGAccuracySeqSplit}Average accuracy of the amplitude
    evaluation using Berends-Giele
    recursion for phase space generation with
    sequential splitting. }
  \end{center}
\end{figure}

\begin{figure}[htb]
  \begin{center}
    \leavevmode
    \includegraphics{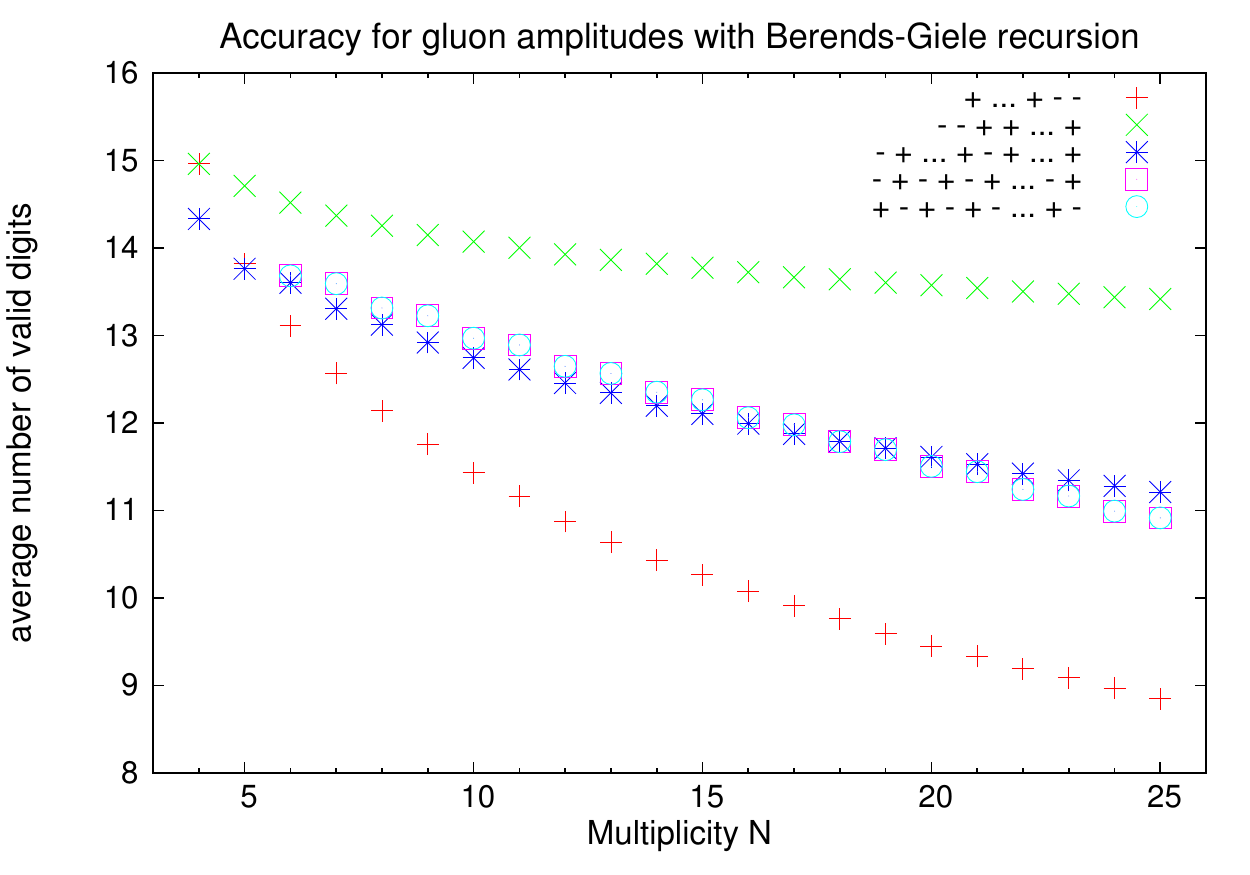}
    \caption{\label{fig:BGAccuracyRambo} Average accuracy of the amplitude
    evaluation using Berends-Giele
    recursion for flat phase space generation.}
  \end{center}
\end{figure}

Understanding the numerical accuracy is crucial for
numerical cross section evaluations. In cases where analytic results
are available it is possible to assess the accuracy of purely numerical
approaches by comparing with analytic results. However the numerical
evaluation of analytic formulae may also be affected by numerical
instabilities. Furthermore a reliable method is also required for
situations where no analytic results are available. In
\cite{Badger:2010nx} the so-called scaling test was proposed. When applying
the scaling test the scattering amplitudes are calculated twice for a given
phase space point: for each phase space point the scattering
amplitudes are calculated for the given momentum configuration. The
evaluation is then repeated for a re-scaled set of momenta. Since the
corresponding effective operators are not renormalized no anomalous
dimension appears. The two evaluations are thus related by their naive mass
dimension:
\begin{equation}
  \label{eq:ScalingTest}
  A_n( p_1,p_2,\ldots,p_n) = x^{n-4} A_n(x p_1,x p_2,\ldots,x p_n).
\end{equation}
As was pointed out in \cite{Badger:2010nx} using a value for $x$ which
is not a power of 2 will lead to a different mantissa in the floating
point representation and thus to different numerics. The method thus
allows to assess the size of rounding errors. To estimate the
numerical uncertainties we have applied the scaling test for a large number of phase
space points. As a measure for the uncertainty we
have evaluated for each phase space point the quantity $\delta$:
\begin{equation}
  \delta = \log_{10}\left(2\left|\frac{ A_1 - A_2}{A_1 + A_2}\right|\right),
\end{equation}
where $A_1$ denotes the result of the amplitude evaluation for
unscaled momenta while $A_2$ is calculated from \Eqn{eq:ScalingTest}.
The quantity $|\delta|$ gives a measure for the valid digits in the
evaluation, i.e. a value of $|\delta| = 3$ would mean that we expect
$\sim 3$ digits to
be correct. As an example we show in \figref{fig:AccPlotSample}
results for the 25 gluon amplitude.
In case of the Berends-Giele recursion the alternating helicity
configuration $+-+-+\ldots$ is evaluated. The remaining three
histograms show results using analytic formulae for MHV, NMHV, and
NNMHV amplitudes. The phase space points are generated using a
sequential splitting algorithm as described in
\cite{BycklingKajantie}. This algorithm does not produce a flat
distribution in phase space. In fact collinear configurations are
preferred.  We note that we always use a couple of default
cuts based on the JADE jet algorithm to avoid singular regions in
the phase space.
In particular we require $(2 p_i\cdot p_j)/s >10^{-10}$.
 We emphasize that as a consequence of the sequential splitting algorithm 
collinear configurations will dominate for multiplicities greater than 15, 
e.g. for $N=20$ almost all phase space points have a collinearity of $10^{-10}$.
It follows from \figref{fig:AccPlotSample} that most of the phase space
points are evaluated with a precision better than 5 valid
digits---largely sufficient for any practical application at hadron colliders.
Since we are mainly interested in a comparison between the purely
numerical approach and the usage of analytic formulae for different parton
multiplicities we have calculated
an average accuracy for different parton multiplicities and different
helicity configurations. The result is shown in
\figref{fig:AvgAccGluons}.
First of all we observe that in general the analytic formulae perform
better as far as the accuracy is concerned. Furthermore it can be seen
that for the analytic formulae the accuracy degrades when we move to
the more complex configurations as for example the NNMHV amplitudes.
This has two reasons. First of all the corresponding formulae
are more involved and are thus more difficult to evaluate numerically. 
The second reason is due to numerical cancellations between individual
terms which leads to a loss of accuracy.  This is supported by the
observation that the NNMHV split helicity where no cancellation occurs
is almost as accurate as the MHV formula.
In the worst case the accuracy is only marginally better than what we
observe in the purely numerical case. For the Berends-Giele recurrence
relations we show only one helicity configuration since no helicity
methods are used as mentioned in the previous section. Describing the
gluon polarization using a four-vector the naive expectation would be
that all helicity configurations should perform similar. A more
detailed analysis shows a mild dependence on the helicity
configuration as can be seen in \figref{fig:BGAccuracySeqSplit}.
To assess the dependence of the aforementioned results on the phase
space generation we show in  \figref{fig:BGAccuracyRambo} the average
accuracy for a flat phase space generation obtained by using the
algorithm RAMBO described in \cite{Kleiss:1985gy}. Comparing
\figref{fig:BGAccuracySeqSplit} and  \figref{fig:BGAccuracyRambo}
we observe two important differences: First of all the dependence of
the average accuracy on the helicity configuration is now more
pronounced than in the case where collinear phase space configurations
were preferred. Second we observe that in particular cases the
accuracy can compete with the numerical evaluation using  analytic formulae.
Our understanding of the observed pattern is the following: Particular
sub-amplitudes or even entire amplitudes may vanish due to ``helicity
conservation''. In the numerical approach this `zero' is `calculated'
from a combination of individual non-zero contributions.  Depending on
the helicity configuration the cancellation may appear earlier or
later in the recursion affecting through accumulated rounding errors
the accuracy of the final result --- leading to the observed helicity
dependence of the average accuracy. In case that collinear phase space
configurations are preferred a second effect becomes important: It is
well known that scattering amplitudes in pure gauge theory show only
square root singularities ($1/\sqrt{p_ip_j}$) for collinear phase
space configurations ($p_i||p_j$). On the other hand individual
Feynman diagrams show a more singular behavior. In the numerical
approach the square root behavior is obtained through a numerical
cancellation of the leading behavior. It is obvious that this leads
to a loss of accuracy. In the extreme case for highly collinear
configurations the 15 digits precision is no longer sufficient to
evaluate a meaningful result.  Using phase space configurations which
tend to be collinear the second effect will dominate the average
accuracy. As a consequence we observe in
\figref{fig:BGAccuracySeqSplit} only a mild dependence of the average
accuracy for different helicity configurations together with an
equally `bad' overall accuracy. (One should keep in mind at this point
that for all practical applications in collider phenomenology 8
significant digits are largely sufficient.) As far as the analytic
formulae are concerned we observe an opposite effect: In case of a
flat phase space generation no particular cancellation in the analytic
formulae is present. As a consequence we observe a very high average
accuracy close to the maximum of about 15 digits as one would have
expected. However studying collinear configurations leads also in the
analytic formulae to cancellations between individual terms. As a
consequence the average accuracy degrades in that case.

One could argue that a flat phase space generation would be more
appropriate to investigate the average accuracy. However we believe
that for practical applications the average accuracy evaluated in that
way would be less meaningful. In phenomenological applications the
cross sections will get important contributions from collinear
configurations. Using Monte
Carlo methods for the cross section evaluation collinear
events will thus dominate the total result. In an ideal situation
this would be taken into account through the phase space integrator by
preferring collinear configurations. This reasoning is also supported
by the empirical observation that using RAMBO for the cross section
evaluation usually  leads to a poor
performance of the Monte Carlo phase space integration in terms of
computational effort and achieved integration accuracy.

\begin{figure}[htb]
  \begin{center}
    \leavevmode
    \includegraphics[width=14cm]{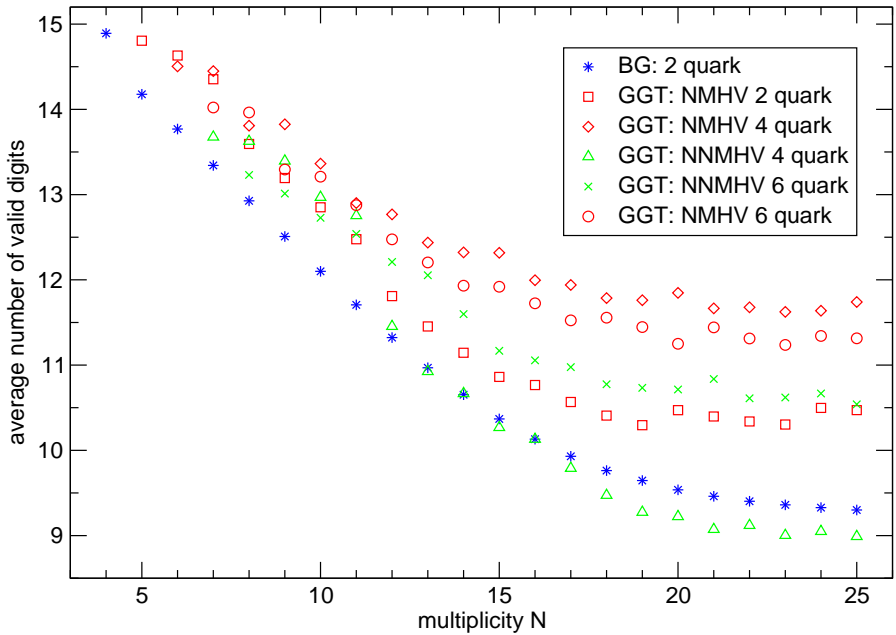}%[angle=-90,width=14cm]{accuracy_quarks}
    \caption{\label{fig:QuarkAccuracy}
      Average accuracy for amplitudes involving quarks.
       Phase-space generation by sequential splitting.}
  \end{center}
\end{figure}

For completeness we analyzed also the average accuracy for amplitudes
involving massless quarks. The result is shown in \figref{fig:QuarkAccuracy}.
Again we have used a phase space generation preferring collinear
events. As far as the numerical approach is concerned the result looks
similar to the pure gluon case. This is just a consequence of the
basic fact that the recurrence relation is very similar apart from the
spin dependence. Since some vertices do not exist in the quark case the
mixed amplitudes contain less terms and are slightly more precise.
Concerning the analytic formulae we observe that the accuracy is not
as good as in the pure gluon case. Our naive understanding is again that the
corresponding formulae are more involved requiring more floating point
evaluations and leaving more room for (unwanted) cancellations in the
case of collinear phase space configurations.

\section{Conclusions}
QCD tree amplitudes are of great interest. The detailed analysis of
their analytic structure may lead to a more profound understanding of
SU(N) gauge theories exposing further symmetries undiscovered so far.
Concerning phenomenological applications tree amplitudes represent an
important input for cross section evaluations in Born approximation
and beyond. In this work we have analyzed two different approaches to
evaluate tree amplitudes. We have compared the numerical performance
of a purely numerical approach based on the Berends-Giele recursion
with the numerical evaluation of analytic formulae. In detail we find
that MHV and NMHV amplitudes are most efficiently calculated using
analytic formulae.  For NNMHV amplitudes and beyond we find the purely
numerical approach more efficient. We have also investigated the
numerical accuracy. In general the numerical accuracy of the analytic
formulae (evaluated numerically) is superior compared to the purely
numerical approach.  However we find that close to exceptional phase
space configuration (soft/collinear configurations) analytic
formulae suffer also from rounding errors. In both approaches we find even
for large multiplicities an average accuracy of at least 9
digits---sufficient for phenomenological applications.

\subsection*{Acknowledgments}
This work is supported in part by the Deutsche Forschungsgemeinschaft
through the Transregional Collaborative Research Centre SFB-TR9
``Computational Particle Physics''and the Research Training Group
(GK1504) "Mass, Spectrum, Symmetry, Particle Physics in the Era of the
Large Hadron Collider" GK1504. In addition we greatfully acknowledge support
from the Helmholtz Alliance {\it ``Physics at the Terascale''}
contract VH-HA-101 and the Volkswagen Foundation.
%%%%%%%%%%%%%%%%%%%%%%%%%%%%%%%%%%%%%%%%%%%%%%%%%%%%%%%%%%%%%%%%%%%%%%%%%%%%%%%%
%%%%%%%%%%%%%%%%%%%%%%%%%%%%%%%%%%%%%%%%%%%%%%%%%%%%%%%%%%%%%%%%%%%%%%%%%%%%%%%%

%%%%%%%%%%%%%%%%%%%%%%%%%%%%%%%%%%%%%%%%%%%%%%%%%%%%%%%%%%%%%%%%%%%%%%%%%%%%%%%%
%%%%%%%%%%%%%%%%%%%%%%%%%%%%%%%%%%%%%%%%%%%%%%%%%%%%%%%%%%%%%%%%%%%%%%%%%%%%%%%%
\bibliographystyle{nb}
\bibliography{botany}

\end{document}